\documentclass[11pt,a4paper]{article}
\usepackage{jinstpub}
\usepackage{rotating}
\usepackage{graphicx}
\usepackage{upgreek}
\usepackage{caption}
\usepackage{subcaption}
\usepackage{array}
\usepackage{multirow}
\usepackage[utf8]{inputenc}

\begin{document}


\title{Sub-Microsecond X-Ray Imaging Using Hole-Collecting Schottky type CdTe with Charge-Integrating Pixel Array Detectors}

\author[a,b]{Julian Becker,}
\author[a]{Mark W. Tate,}
\author[a]{Katherine S. Shanks,}
\author[a]{Hugh T. Philipp,}
\author[a,b]{Joel T. Weiss,}
\author[a]{Prafull Purohit,}
\author[b]{Darol Chamberlain,}
\author[a,b,c,1]{and Sol M. Gruner\note{Corresponding author}}

\affiliation[a]{Laboratory of Atomic and Solid State Physics, Cornell University, Ithaca, NY 14853,USA}
\affiliation[b]{Cornell High Energy Synchrotron Source (CHESS), Cornell University, Ithaca, NY 14853, USA}
\affiliation[c]{Kavli Institute at Cornell for Nanoscale Science, Cornell University, Ithaca, NY 14853, USA}
\emailAdd{smg26@cornell.edu}


\abstract{CdTe is increasingly being used as the x-ray sensing material in imaging pixel array detectors for x-rays, generally above 20 keV, where silicon sensors become unacceptably transparent. Unfortunately CdTe suffers from polarization, which can alter the response of the material over time and with accumulated dose. Most prior studies used long integration times or CdTe that was not of the hole-collecting Schottky type. We investigated the temporal response of hole-collecting Schottky type CdTe sensors on timescales ranging from tens of nanoseconds to several seconds. We found that the material shows signal persistence on the timescale of hundreds of milliseconds attributed to the detrapping of a shallow trap, and additional persistence on sub-microsecond timescales after polarization. The results show that this type of CdTe can be used for time resolved studies down to approximately 100~ns. However quantitative interpretation of the signal requires careful attention to bias voltages, polarization and exposure history.}

\maketitle

\section{Introduction}


Pixel Array Detectors (PADs) have been used for x-ray science since the late 90s \cite{I1}, and have arguably changed the detector landscape at synchrotron sources. One of the core ideas of PADs is to separate the sensor, which detects radiation, from the signal processing by introducing two distinct layers, keeping the Application Specific Integrated Circuit (ASIC) apart from the sensitive volume. This approach has the advantage that sensor material and read out can be optimized independently. To date silicon is the most commonly used sensor material. Silicon sensors are readily available at low cost with excellent quality, however its stopping power for x-rays above approximately 20~keV is diminishing, which greatly reduces the probability to detect individual high energy photons.

\begin{figure*}[tb]
  \centering
  \includegraphics[width=0.8\textwidth]{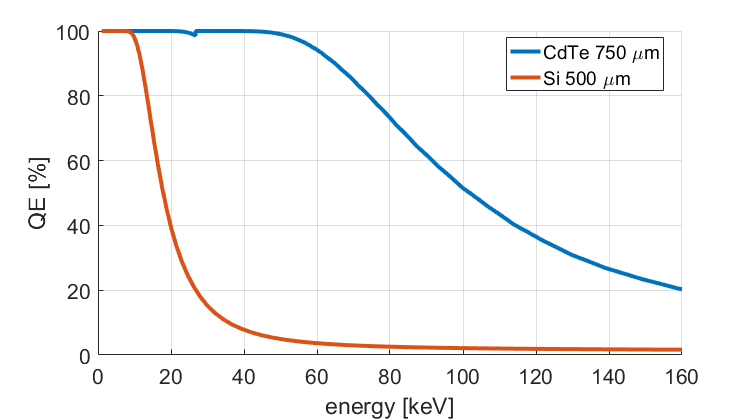}
  \caption{Quantum efficiency of 750~$\upmu$m CdTe and 500~$\upmu$m silicon sensors. }
  \label{QE}
\end{figure*}

At energies above 20~keV other approaches exist \cite{detectors}, each having its own advantages and disadvantages. Cadmium Telluride (CdTe) is considered to be one of the most promising sensor materials for high energy application. Its quality and availability has increased significantly \cite{R1, R2} and scientific x-ray imagers with CdTe sensors are available from several commercial vendors. Cornell University has developed multiple detector systems in the past (see below for details). For the present study we replaced the standard 500~$\upmu$m thick silicon sensors with 750~$\upmu$m thick CdTe sensors in order to enable experiments at higher x-ray energies \cite{static}.  Using CdTe sensors increases the quantum efficiency dramatically above 15~keV compared to the standard silicon sensors, as shown in Figure \ref{QE}. 


Characterization of the static behavior of the resultant CdTe PADs has been previously published~\cite{static}. In this publication we present studies of the dynamic response of the CdTe PADs. We performed the characterization with a range of tests using lab sources as well as synchrotron radiation from the Cornell High Energy Synchrotron Source (CHESS) and the Advanced Photon Source (APS) at Argonne National Lab.

\subsection{Detector systems}
Cornell University has developed several ASICs for charge integrating detectors using silicon sensors. ASICs developed for the Keck PAD detector \cite{K1,K2,K3,K4} and the MM-PAD detector \cite{M1,M2,M3} have been hybridized with hole-collecting Schottky type CdTe sensors instead of silicon sensors for this study. Both detectors (and their ASICs) feature 128~$\times$~128 pixels of (150~$\upmu$m)$^2$ per chip. 

The MM-PAD ASIC uses an in-pixel charge removal circuit to extend its dynamic range by many orders of magnitude, maintaining single x-ray sensitivity \cite{M4, mm_gain} while achieving a dynamic range of $>$~4$\times$10$^7$ 8~keV x-rays/pixel/frame in addition to framing at $>$~1 kHz. This has proven to be very useful for coherent x-ray imaging \cite{M5}.

The Keck PAD ASIC features an array of configurable capacitors in each pixel that can be employed in multiple ways. The burst mode operation is most relevant for this study, as it allows the rapid acquisition of up to 8 distinct frames at up to 10~MHz frame rate onto an in-pixel capacitor array and readout of these stored frames at approximately 1~kHz frame rate.


\subsection{Relevant material properties of CdTe}

The CdTe material used for the sensor investigated here is 750~$\upmu$m thick and was produced by Acrorad Co., Ltd., Japan \cite{acrorad}. The sensors are In/Pt Schottky type with approximate layer thicknesses indicated in Table~\ref{layers}. The ohmic (Pt) contact was pixellated and connected to the ASIC by bump bonds to collect holes. Fabricating the sensors with ohmic contacts (Pt on both sides) or pixellating the other contact was considered, but abandoned due to our requirements of hole collection and low leakage currents. Bump deposition and flip-chip bonding was performed by Oy Ajat Ltd., Finland \cite{ayat}. Placement and gluing of the modules onto heat sinks was performed in-house and wire-bonding of the ASIC was done by Majelac Technologies LLC, USA \cite{majelac}. The HV wire used to bias the sensor was attached in-house using conductive silver paint (Leitsilber 200 from \cite{ted_pella}). All adhesives were cured at room temperature.

\begin{table}
\centering
\begin{tabular}{ |l|c|c|c|c|c|c|c|c| }
  \hline
  \textbf{\emph{Material}}	& Ti & In & CdTe & Pt & Au & Ni & Au & AlN	\\
\hline
   \textbf{\emph{Thickness}} 	&  30~nm & 300~nm & 750~$\upmu$m & 50~nm & 30~nm  & 50~nm & 80~nm & 100~nm \\
  \hline
\end{tabular}
\caption{Approximate layer thicknesses of the CdTe sensor. X-rays enter the sensor from the titanium passivated indium electrode.}
\label{layers}
\end{table}

Using standard lab sources and low dose rate experiments at synchrotron light sources, the detector systems used in this study have been characterized previously \cite{static}, with the expected finding that the CdTe systems polarize after a given dose and/or time. Polarization manifests as changes of more than one detector property. Typical effects include reduction of the collected charge compared to the unpolarized state and the degradation of the resolution. These effects have been characterized and reset schemes to remove these effects were developed.

\section{Basic testing}
The measurements presented in this section were performed using an MM-PAD detector, to make use of the large dynamic range, with a freshly reset\footnote{We followed the procedure outlines in \cite{static}, applying a forward bias of 5~V for 1 minute to reset the diode.}, unpolarized CdTe sensor. All measurements were accompanied by corresponding measurements with a silicon version of the detector next to the version with CdTe sensor for comparison and to exclude any effects from the x-ray tube. The overall gain of the MM-PAD system with CdTe sensor is approximately 0.63~keV per analog to digital unit (ADU) and varies with bias voltage and operating temperature as outlined in \cite{static}.

MM-PAD detectors are currently limited to a maximum bias voltage of 300~V by the supporting electronics, which were designed for use with silicon sensors at $\leq$~200~V.

\subsection{Effects of changing the bias voltage}
Resetting the sensor after polarization involves a change of the detector bias voltage. To measure the sensor response to bias voltage changes we take frames at 100~Hz (9.14~ms integration time per image plus 0.86~ms for readout) for a total duration of 10 seconds. Data acquisition starts close to the end of the reset phase (-5 V applied to HV side of the sensor). Applying the bias to the sensor is done in a gradual manner, utilizing the current limiting functionality of the employed Keithley 2410 source meter. This results in a smooth ramping of the high voltage with a maximum current of 105~$\upmu$A. The exact duration of ramping the bias voltage depends on temperature and on the target bias and is about 1~s for 100~V, 1.7~s for 200~V and 2.2~s for 300~V final bias voltage.

\begin{figure*}[tb!]
  \centering
  \begin{subfigure}[t]{0.45\textwidth}
	\includegraphics[width=\textwidth]{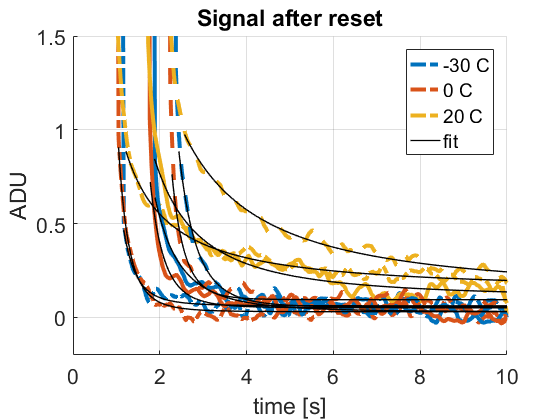}
	\caption{Settling of the sensor after reset and bias ramp.  Data for 100~V, 200~V and 300~V are marked by dash-dotted, solid and dashed lines, respectively. Sensor temperature is indicated by  color. Black lines show fits to the corresponding curves.}
	\label{res_a}
  \end{subfigure}
\quad
  \begin{subfigure}[t]{0.45\textwidth}
	\includegraphics[width=\textwidth]{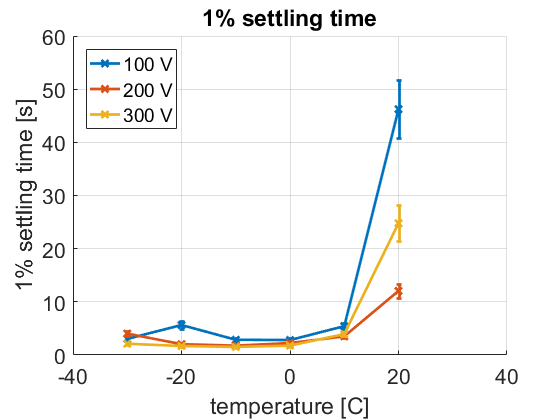}
	\caption{Settling time of the sensor as a function of final bias voltage and temperature. The settling time was calculated as the time required for the signal to settle to 1\% of the initial value.}
	\label{res_b}
  \end{subfigure}
  
  \caption{Response transients and settling time after bias ramps to different voltages at different sensor temperatures.}
  \label{reset}
\end{figure*}

Figure \ref{reset} shows the settling of the sensor response after the reset and the associated voltage ramp. Each data point is an average of the central part of each image, i.e., a rim containing the 15 outermost rows and columns was excluded. The displayed ADUs are offset corrected with pedestal values determined pre-reset. 
Figure \ref{res_a} shows a different starting point of the settling depending on the duration of the bias ramp. All settling curves are generally of similar shape and can be fit to a power law decay. Although not obvious for all displayed data, the sensor eventually settles to the pre-reset values.

As motivated in \cite{static}, we have chosen 0~C as the operating point of our detector. 
Figure \ref{res_b} shows the response transients and settling time for different ramp voltages and sensor temperatures. Since the settling signal could be described by a power law, we defined the settling time as the time required for the signal to decay to 1\% of a given starting point. A general trend of faster settling at lower sensor temperatures and higher voltages is observed. The sensor response settles within 10~s at low temperatures, at 20~C the settling time increases to about one minute at 100~V bias. The reason of this huge increase in settling time and its apparent inconsistency as a function of voltage remains unresolved at this point, but could be related to the increased dark current of the sensor at this temperature.

The observed settling times for the investigated Schottky material are much faster than the commonly suggested settling wait of 30~--~60~minutes after applying a bias voltage to ohmic material \cite{lambda}.

\subsection{Persistence on millisecond timescales}

\begin{figure*}[tb!]
  \centering
  \begin{subfigure}[t]{0.45\textwidth}
	\includegraphics[width=\textwidth]{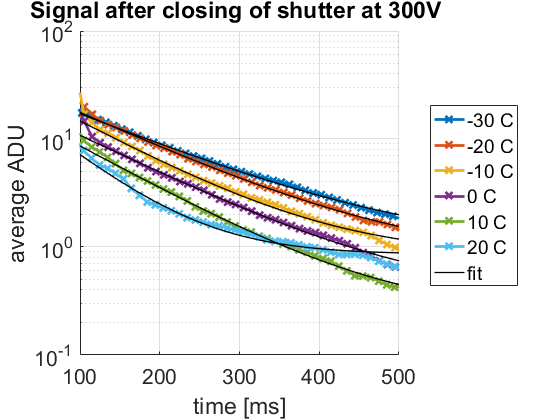}
	\caption{Signal remaining as a function of temperature and corresponding exponential decay fits (black lines) at 300~V bias and 0.858 mA x-ray tube current. Decay curves for 0.444~mA tube current are similar in amplitude and decay constant.}
	\label{p_a}
  \end{subfigure}
\quad
  \begin{subfigure}[t]{0.45\textwidth}
	\includegraphics[width=\textwidth]{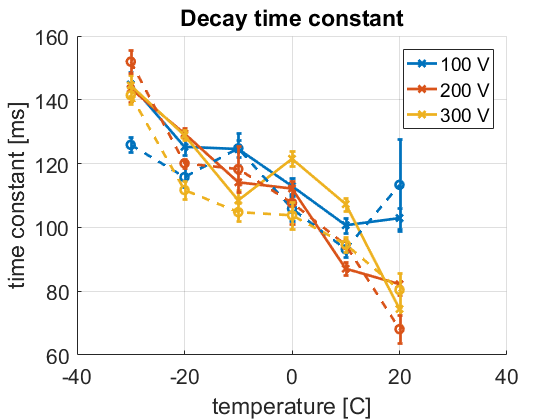}
	\caption{Decay constant of the persistence as a function of bias voltage and temperature. Solid and dashed lines indicate data taken with 0.858~mA and 0.444~mA x-ray tube current, respectively.}
	\label{p_b}
  \end{subfigure}
  
  \caption{Persistence of the signal in response to x-rays after closing the shutter.}
  \label{per}
\end{figure*}

The persistence of the sensor signal after x-ray illumination was studied with a freshly reset, unpolarized sensor and a lab x-ray tube. We define persistence as a signal above background that is recorded immediately after, but in absence of, x-ray illumination. 
We used a 50~W silver (Ag) anode x-ray tube (Trufocus TCM-5000M) at an acceleration voltage of 47~kV in close proximity to the detector (about 5 cm). The spectrum was filtered by 1~mm aluminum and the dose rate on the detector was 3.3~Gy/min and 2.0~Gy/min at 0.858~mA and 0.444~mA tube current, respectively. The tube is mounted to a shutter assembly that opens or closes by mechanically moving a tungsten disk into the beam. The shutter operates in less than 10 ms, however mechanical vibrations cause systematic effects for about 100~ms after the shutter closes and this time was therefore excluded from the data analysis.

For comparison, we measured the response of a silicon diode after closing the shutter and could not observe a signal above background.

At an x-ray tube current of 0.858~mA the response of the CdTe diode was measured for illumination times of 1~s, 10~s and 100~s. We obtained identical decay curves in each case, leading to the conclusion that the cause of the persistence saturates, possibly by reaching an equilibrium state between trapping and detrapping of the involved traps.

The decay of the persistent signal, $I(t)$, was assumed to be casued by a single shallow trap level and fit to a simple exponential decay:
\begin{equation}
I(t) = A e^{-\frac{t}{\tau}} + b 
\label{decay}
\end{equation}

where $I(t)$ is the sensor response after the shutter closes, $A$ is an amplitude constant, $t$ is the time after closing of the shutter, $\tau$ is the decay time constant and $b$ is an offset introduced by the polarization of the sensor.

Figure \ref{per} shows selected decay transients and the decay time constants obtained from the fits to all measured data ranging from approximately 150~ms at -30~C to about 90~ms at +20~C. The decay constant does not clearly indicate any dependence on applied bias or x-ray tube current.


\subsection{Defect characteristics}

\begin{figure*}[tb!]
  \centering
	\includegraphics[width=1.0\textwidth]{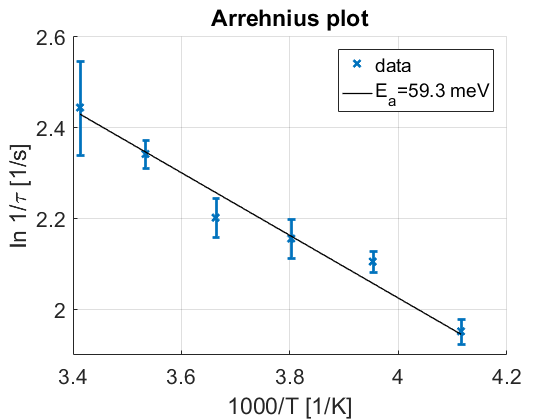}
	\caption{Logarithm of the inverse decay constant as a function of the inverse temperature. The data are consistent with an activation energy of 59.3~meV.}
	\label{t_a}

\end{figure*}

The temperature dependence observed for the decay constant, $\tau$, follows the Arrhenius equation, $1/\tau=C \exp\left(-E_a / k_b T\right)$, with $C$ being an arbitrary constant, $k_b$ Boltzman's constant and $T$ the absolute temperature. The trap activation energy, $E_a$, is determined from a fit to the data and has a value of 59.3~$\pm$~7.2~meV, as shown in Figure \ref{t_a}. A similar defect level has been observed in the past in similar CdTe material \cite{trap2}.


We observe that the absolute magnitude of the persistence signal, defined by the amplitude constant, $A$ in Equation \ref{decay}, is independent of illumination time (for times $\geq$ 1~s), tube current\footnote{Since the photocurrent, $I_0$, is proportional to the tube current the relative magnitude of the persistence, $A/I_0$, decreases as $I_0$ increases.} and temperature, but depends on the applied voltage. This indicates that the total persistence signal is limited by the number of filled traps, with the equilibrium level of trapping and detrapping being a function of the applied voltage. 

Integrating the fit to the measurement data we are also able to estimate how many traps were filled at the moment the shutter closes. This value is a function of the applied bias and is 1.04~$\times$~10$^9$~cm$^{-3}$ $\pm$ 8\%, 1.78~$\times$~10$^9$~cm$^{-3}$ $\pm$ 6\% and 2.16~$\times$~10$^9$~cm$^{-3}$ $\pm$ 10\% at 100, 200 and 300~V, respectively. For this we assumed that each trap is only trapping a single hole, and used the average deposited energy per created electron-hole pair, 4.43~eV/pair, a material constant. The a total charge, released over the whole decay time, is equivalent to approximately 77~--~161~keV deposited energy per pixel.

The determined trap densities are unusually low for a bulk defect, as is the fact that the defect level does not seem to depend on the applied bias voltage. Thus we cannot exclude that other defects, e.g., interface states at the contacts, cause the observed behavior. 


\section{Charge transport measurements using a synchrotron source}
\begin{figure*}[tb]
  \centering
  \includegraphics[width=1.0\textwidth]{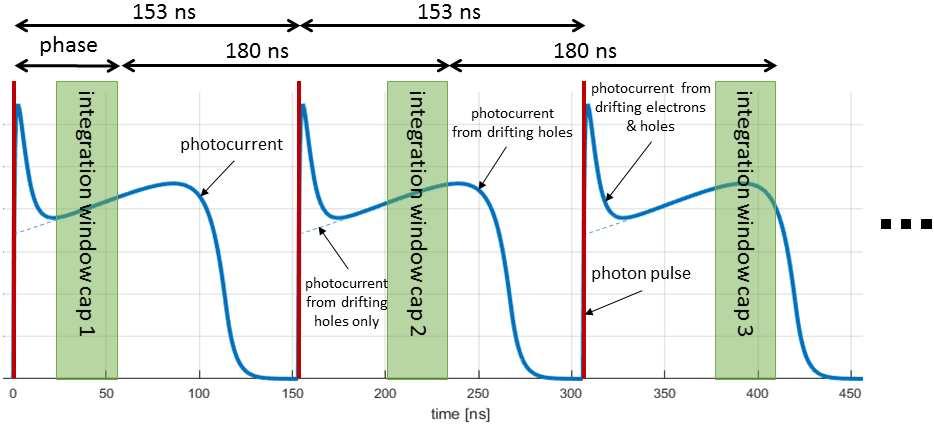}
  \caption{Schematic of the photocurrent for 108~keV photons and timing properties for the APS experiment. The detector was triggered with an adjustable phase delay with respect to the x-ray pulses arriving every 153~ns. The red lines depict the x-ray pulses, each derived from a single bunch of electrons in the APS ring. Timing is shown for the first 3 of 8 frames in a single burst.}
  \label{APS_timing}
\end{figure*}

Capitalizing on the fast imaging capabilities of the Keck PAD system, we studied the charge transport in the CdTe sensor material with short but intense photon pulses at the Dynamic Compression Sector beamline 35IDE at the Advanced Photon Source at Argonne National Laboratory. These data were taken with a freshly reset, unpolarized sensor. The overall gain of the Keck PAD system with CdTe sensor is approximately 5.5~keV/ADU and varies with bias voltage and operating temperature as outlined in \cite{static}.

The storage ring was operated in 24 bunch mode with bunches evenly spaced 153~ns apart. The beamline was operated in unfocused pink beam mode with the undulator gap adjusted to produce a 7.1~keV x-ray fundamental. X-ray harmonics of this fundamental energy were produced to >~120~keV. A rotating chopper produced an isolated 20~$\upmu$s window of x-rays to limit the total exposure of the sensor.

X-ray energies were separated utilizing diffraction from a 0.5~mm thick sheet of copper. Crystal grains in this sheet were roll aligned, producing a series of spatially isolated <200> diffraction spots, each at a different energy. Data presented here are taken from diffraction spots at 42, 69, and 108~keV.




The detector was operating at 0~C for the experiment. The start of a burst sequence was synchronized to the ring frequency with an estimated time jitter of less than 3~ns and the trigger was set up such that the phase delay between the detector and the photon pulses could be freely adjusted. As illustrated in Figure \ref{APS_timing}, burst framing was utilized to sample successive portions of single pulse photocurrent transients, which ultimately allows the reconstruction of these transients. Each of the eight frames in a burst sample has the same length of the integration window ($\approx$~35~ns). Although the photon pulses themselves are extremely short ($\ll$~1~ns typically), drift and diffusion processes in the sensor material slow the charge collection process down to $\approx$~100~ns, depending on the applied bias.

The results obtained are compared to simulations done with a custom MATLAB code \cite{static, sim1, sim2}, which was further modified to account for an effective RC delay of 5~ns (slew rate limitation of the pixel preamplifier in the ASIC) and the size of the beam spot on the detector.

\begin{table}
\centering
\begin{tabular}{ |l|c|l| }
  \hline
  \textbf{\emph{parameter}}		& \textbf{\emph{value}} 		& \textbf{\emph{unit}}\\
  \hline
  CdTe sensor thickness	& 750							& $\upmu$m \\
  pixel size			& 150							& $\upmu$m \\
  effective integration window & 35 						& ns \\
  x-ray attenuation length & 102~/~383~/~1257				& $\upmu$m \\
  beam width on detector	& 5.77~/~1.23~/~0.86				& mm \\
  mobility (electron/hole)	& 1100~/~82.6					& cm$^2$/Vs \\
  trapping time (electrons and holes)	& 3					 & $\upmu$s \\
  space charge density $N_{eff}$	& 0.03~/~3.11~/~4.90				& 10$^{11}$ cm$^{-3}$ \\
  \hline
\end{tabular}
\caption{Device parameters and material properties used in the numerical simulations. Values separated by `/' were used for the simulations of the 42~keV, 69~keV, and 108~keV cases, respectively.}
\label{sim_pars}
\end{table}

\begin{figure*}[tb!]
  \centering
  \begin{subfigure}[t]{0.45\textwidth}
	\includegraphics[width=\textwidth]{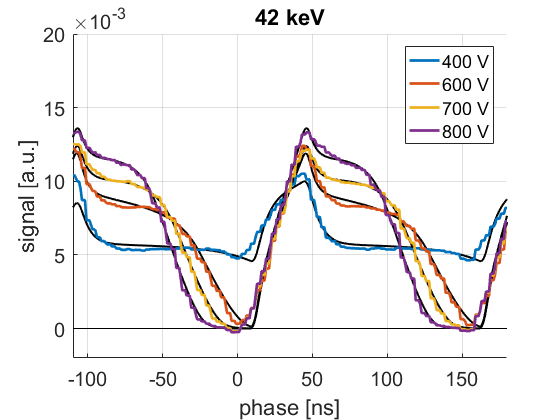}
	\caption{Response for 42~keV beam energy. The curve for 400~V shows pulse pile up as the charge collection time exceeds 153~ns.}
	\label{p_42}
  \end{subfigure}  
  \quad
  \begin{subfigure}[t]{0.45\textwidth}
	\includegraphics[width=\textwidth]{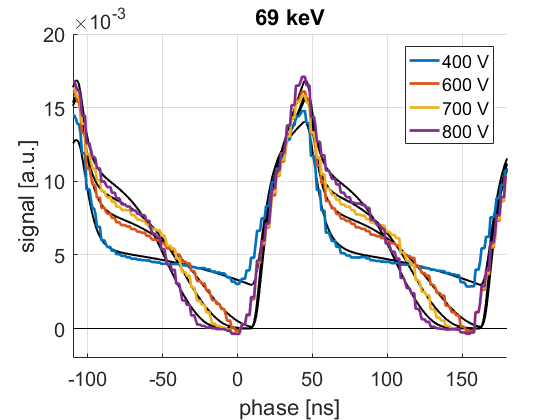}
	\caption{Response for 69~keV beam energy. The curve for 400~V shows pulse pile up as the charge collection time exceeds 153~ns.}
	\label{p_69}
  \end{subfigure}  

\smallskip

   \begin{subfigure}[t]{0.45\textwidth}
	\includegraphics[width=\textwidth]{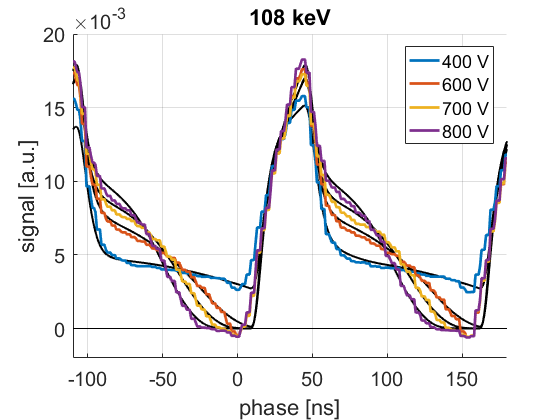}
	\caption{Response for 108~keV beam energy. The curve for 400~V shows pulse pile up as the charge collection time exceeds 153~ns.}
	\label{p_108}
  \end{subfigure}  
  \quad
  \begin{subfigure}[t]{0.45\textwidth}
	\includegraphics[width=\textwidth]{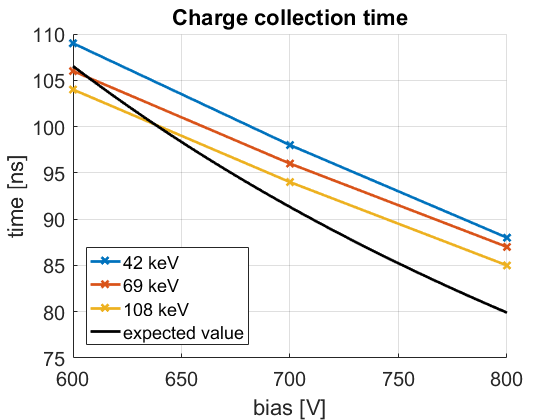}
	\caption{Charge collection time (5\% to 95\%) and expected charge collection time for $\mu_h= 88$~cm$^2$/Vs. The charge collection time for 400~V exceeds 153~ns for all energies.}
	\label{cct}
  \end{subfigure}
  \caption{Response of the Keck PAD detector to short pulses of different energies as a function of applied bias. Black lines indicate the expected signal from a numerical simulation of the experiment.}
  \label{phasing}
\end{figure*}

Table \ref{sim_pars} summarizes the material properties and device parameters used for the simulations, and Figure \ref{phasing} shows that the modeling and the measured response are in good agreement. The results are consistent with no effective space charge density\footnote{Effective space charge density is defined here as the difference of all immobile positive charges and all immobile negative charges within a pixel volume. Examples of immobile charges include certain defects and charged trap states.} in the case of 42~keV and a positive effective space charge density of 3~$\times$~10$^{11}$~cm$^{-3}$ and 5~$\times$~10$^{11}$~cm$^{-3}$ for the 69 and 108~keV case, respectively. We find that our derived space charge density is the same order of magnitude as observed by others (10$^{11}$--10$^{12}$~cm$^{-3}$ \cite{TCT1}, 2.37--8.69~$\times$~10$^{11}$~cm$^{-3}$ \cite{TCT2}) for ohmic material from the same manufacturer.

The charge collection time is dominated by the hole mobility. For the best agreement between measurements and simulations a mobility value of 82.6 cm$^2$/Vs was used. This value is comparable to literature values of 88 cm$^2$/Vs \cite{mob1} and 72~$\pm$~11 cm$^2$/Vs \cite{TCT2} for material from the same manufacturer at room temperature.

For investigations of fast processes on microsecond time scales it is mandatory that the charge collection time of the sensor is short compared to the timescale of interest. Figure \ref{cct} shows the measured charge collection time\footnote{Measured charge collection time is defined here as the time between the points where 5\% and 95\% of the total charge is collected.} as a function of voltage compared to the charge collection time expected from literature values. For voltages above 600~V the charge collection time is below 120~ns. Although the measured values are slightly larger than the expected values, they are small enough to allow imaging without pileup effects in the sensor at repetition rates up to approximately 8~MHz.

\newpage
\section{Polarization and persistence after high rate illumination}

The fast framing capability of the Keck PAD system was used to study the sensor response to short, intense pulses of photons, as typically encountered during synchrotron experiments. The following experimental data were obtained at the A1 beamline of the Cornell High Energy Synchrotron Source (CHESS). 

The CHESS beamline A1 receives x-rays produced by electrons passing through one of two 1.5 m CHESS Compact Undulators \cite{c1, c2}  installed in the west flare of Cornell Electron Storage Ring (CESR). A1 is equipped with a switchable diamond monochromator that enables selection of either 19.9~keV or 32.6~keV x-ray energy by using either the diamond <111> or the diamond <220> Bragg condition. 

Mirrors that would reduce the harmonic content of the beam were not used. All beam defining slits were fully opened. The beam was attenuated by aluminum absorbers of 8.4 and 21.2~mm thickness to increase the relative content of higher harmonics at 59.7~keV (three times fundamental at 19.9~keV) and 65.2~keV (two times fundamental at 32.6~keV), respectively.

A flux estimate was determined by using an ion chamber in front of the experimental setup in addition to the detector signal. By using the two readings and the known attenuation at each energy, the flux of both the fundamental and the harmonics could be determined at the same time. The results summarized in Table \ref{flux_a1} represent typical values; actual values are modulated by the beam current in the ring, which continuously decays between injections. 

At the time of the experiments the electrons in the storage ring were arranged in 3 `trains' of 16 bunches each, where each bunch is of $\ll$~1~ns duration. Inside each train the electron bunches are separated by 4~ns each, resulting in a total duration of 60~ns for each train. The time between the first bunch of the first train and the first bunch of the second train is 252~ns and is identical to the time between the first bunches of second and third train. From the beamline's point of view this pattern translates to photon pulses that are repeated every 2.56~$\upmu$s when the electrons have completed another orbit of the storage ring. 

\begin{table}
\centering
\begin{tabular}{ |l|c|c|c|c| }
  \hline
  & \multicolumn{2}{c|}{19.9 keV} & \multicolumn{2}{c|}{32.6 keV}\\
  			& before att. 			& after att.  				& before att. 			& after att. \\
  \hline
  fundamental (f)	& 3.67 $\times$ 10$^{11}$	& 6.38 $\times$ 10$^{7}$ 	& 1.27 $\times$ 10$^{11}$ 	& 3.62 $\times$ 10$^{8}$ \\
  harmonic (h) 	& 1.43 $\times$ 10$^{10}$	& 7.27 $\times$ 10$^{9}$ 	& 3.17 $\times$ 10$^{10}$ 	& 6.92 $\times$ 10$^{9}$ \\
   \hline \hline
  ratio (h/f)	 	& 1/26				& 114 					& 1/4				 	& 19 \\
  \hline
\end{tabular}
\caption{Flux estimate in photons per second for the experiment at CHESS beamline A1, in a footprint of about 1~mm~$\times$~1~mm. Aluminum pieces of 8.4 and 21.2~mm thickness were used to attenuate the 19.9~keV and 32.6~keV beam, respectively. The ratio indicates how many harmonic photons are present in the beam for every photon of the fundamental energy.}
\label{flux_a1}
\end{table}

\begin{figure*}[tb]
  \centering
  \includegraphics[width=1.0\textwidth]{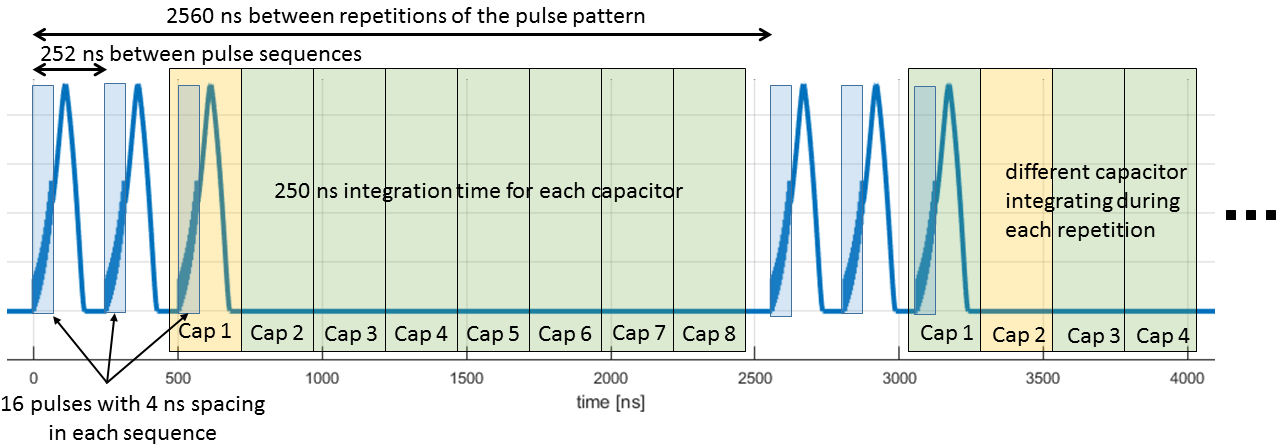}
  \caption{Schematic of the photocurrent for 65.2~keV and the timing properties for the CHESS experiment. The detector was triggered such that the capacitors sampled different times with respect to the photon pulse sequence.}
  \label{CHESS_timing}
\end{figure*}

For the experiment the detector was synchronized to the ring frequency such that the photocurrent caused by the last train in the pulse pattern was recorded, as well as the dark period in between repetitions of the pattern, as sketched in Figure \ref{CHESS_timing}. Within the ASIC, each frame was integrated onto a seperate storage capacitor for 250~ns with the time between the start of each acquisition equal to (2560+250)~ns. This way the time sequence is sampled without a gap within 8 turns of the electron bunches around the storage ring. 
The estimated time jitter between the start of the photon pulse sequence and the integration time was less than 10~ns.

\begin{figure*}[tb!]
  \centering
  \begin{subfigure}[t]{0.45\textwidth}
	\includegraphics[width=\textwidth]{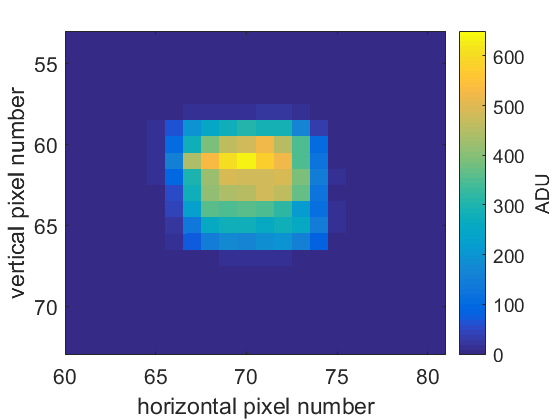}
	\caption{Average signal}
	\label{beam}
  \end{subfigure}
\quad
  \begin{subfigure}[t]{0.45\textwidth}
	\includegraphics[width=\textwidth]{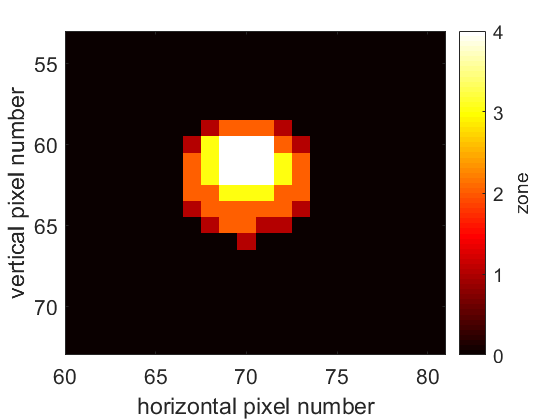}
	\caption{Zones with different dose rates.}
	\label{zone_map}
  \end{subfigure}
  
  \caption{Typical beam profile at the CHESS beamline A1, as seen on the detector at the time of the experiment.}
  \label{zones}
\end{figure*}

This sampling of the time sequence was done for a series of dose steps to study the influence of polarization. Each dose step illuminated the sensor for 50~ms, and a total of 500 sequential steps (25~s total illumination time) were recorded before resetting the detector. Since the photon beam was not slit down or focused it spanned several pixels and showed a notable beam profile, which allowed us to group several pixels into zones with different dose rates as shown in Figure \ref{zones}. The average flux and corresponding dose rate for each zone is shown in Table \ref{dose_rate}.

Dose series were acquired for a range of different temperatures and bias voltages to study the effect of these parameters on the deterioration due to the polarization.

Previous investigations using a Keck PAD system with silicon sensor \cite{K4} determined that the resolution of the time response was limited by the charge collection time of the silicon, some ten nanoseconds, and not the electronic response of the system.


\begin{table}
\centering
\begin{tabular}{ |l|c|c|c|c| }
  \hline
  & \multicolumn{2}{c|}{59.7 keV} & \multicolumn{2}{c|}{65.2 keV}\\
  		& flux 	[ph/mm$^2$/s]		& dose rate [Gy/s]		& flux 	[ph/mm$^2$/s]		& dose rate [Gy/s] \\
  \hline
 Zone 1 	& 1.05 $\times$ 10$^{10}$	& 8.5				& 1.44 $\times$ 10$^{10}$ 	& 12.5 \\
 Zone 2	& 1.43 $\times$ 10$^{10}$	& 11.6			& 1.91 $\times$ 10$^{10}$ 	& 16.6 \\
 Zone 3	& 1.84 $\times$ 10$^{10}$	& 14.9			& 2.37 $\times$ 10$^{10}$	& 20.6\\
 Zone 4	& 2.56 $\times$ 10$^{10}$	& 20.8			& 2.53 $\times$ 10$^{10}$	& 22.0 \\
  \hline
\end{tabular}
\caption{Estimate of the average flux and dose rate for the different zones.}
\label{dose_rate}
\end{table}

\subsection{Mathematical formalism}

We assume that the sensor response to the incoming photon bunch can be described by a prompt photocurrent, $I_p(t)$, and, especially once the sensor is polarized, a delayed release of trapped current, $I_d(t)$. As shown in the previous section, the prompt photocurrent is well described by drift and diffusion simulations. Thus all of the prompt signal is collected in less than 250~ns, the integration time, $t_i$, used in this experiment. Making the ansatz that the delayed photocurrent is predominately caused by the detrapping of charge from a single shallow trap level\footnote{Although decribed by the same equation, this second trap is different from the previously described first trap causing signal persistence on millisecond timescales.}, it can be expressed as:

\begin{equation}
I_{d}\left(t^\prime\right) = I_0 e^{-\frac{t^\prime}{\tau}}
\label{I_d}
\end{equation}

The above is valid for times $t^\prime>0$, where $t^\prime$ is defined as $t^\prime= t - t_{end}$, the time passed since the charge collection process is finished, with $t_{end}$ being the time when the charge collection of the prompt response has finished with respect to the start of the integration window and $I_0$ a constant that is proportional to the total number of trapped charges. 

The total sensor response (total signal charge received) is the sum of both responses, prompt and delayed, with $R_{tot} = P_{tot} + D_{tot}$, and :

\begin{eqnarray}
P_{tot} =& \int_0^{t_{end}} &I_p(t) \; dt 	\\
D_{tot} =& \int_{0}^{\infty} &I_d(t) \; dt \;=\; I_0 \tau
\label{tot}
\end{eqnarray}

Defining $\delta_1$ and $\delta_2$ as the time difference between the first pulses of the first and the third, or the second and the third bunch train, 504~ns and 252~ns, respectively, and assuming that the incoming intensity per pulse is equal in every train, we can express $S(n)$, the signal collected on capacitor $n$, for $n>1$ in the following way:

\begin{eqnarray}
S\left(n\right) &=& c + \int_{t_0}^{t_0+t_i} I_d(t) +  I_d(t+\delta_1) +  I_d(t+\delta_2) \; dt \\
 &=& c + I_0 \tau \left( 1 - e^{-\frac{t_i}{\tau}} \right) \left( 1+ e^{-\frac{\delta_1}{\tau}}+ e^{-\frac{\delta_2}{\tau}} \right) e^{-\frac{t_0}{\tau}} \\
 & = & A B C e^{-\frac{t_0}{\tau}}+ c
\label{S_n}
\end{eqnarray}

We define $c$ as an offset compensating for shifts in the zero level due to polarization of the sensor, $t_0 = t_i (n-1) - t_{end} $ as the time defining the start of the integration window with respect to $t^\prime=0$, $ A=  I_0 \tau = D_{tot}$, $B=\left( 1 - e^{-\frac{t_i}{\tau}} \right)$ a factor accounting for the finite width of the integration window, and $C = \left( 1+ e^{-\frac{\delta_1}{\tau}}+ e^{-\frac{\delta_2}{\tau}} \right)$ a factor ranging from 1 to 3 for values of $\tau$ ranging from zero to infinity which accounts for the presence of 3 trains instead of a single train.

As shown in Figure \ref{fits}, the signal of the delayed release (capacitors 2 - 8) can be described well by the exponential decay of equation \ref{S_n}. The values of $I_0\tau = D_{tot}$ and $c$ can be determined from a fit to the data since the values for $t_i$, $t_{end}$, $\delta_1$, and $\delta_2$ are known.

The intensity recorded by the first capacitor, $S(1)$, is the sum of the prompt response $P_{tot}$, the integral of the delayed response from $t_{end}$ to the end of the integration window, and the delayed contributions from the previous pulses. It is possible to rewrite $S(1)$ and solve for $P_{tot}$ in the following way:

\begin{eqnarray}
S\left(1\right) &=& P_{tot} +  I_0 \tau \left( 1 - e^{-\frac{t_i-t_{end}}{\tau}} \right) +  I_0 \tau \left( 1 - e^{-\frac{t_i}{\tau}} \right) \left( e^{-\frac{\delta_1}{\tau}} + e^{-\frac{\delta_2}{\tau}} \right) + c \\
P_{tot} &=& S\left(1\right) -  I_0 \tau \left[  \left( 1 - e^{-\frac{t_i-t_{end}}{\tau}} \right) + \left( 1 - e^{-\frac{t_i}{\tau}} \right) \left( e^{-\frac{\delta_1}{\tau}} + e^{-\frac{\delta_2}{\tau}} \right) \right] -c
\label{I_0}
\end{eqnarray}

Since $R_{tot}$, the response of CdTe to photons, decreases as polarization sets in, it is convenient to define the fraction of charge that is delayed, $F$, which can be expressed in the following way:

\begin{eqnarray}
F &=& \frac{D_{tot}}{R_{tot}} \\
  &=& \frac{D_{tot}}{ P_{tot} + D_{tot}} \\
  &=& \frac{I_0 \tau}{ P_{tot} +I_0 \tau} \\
  &=& \frac{I_0 \tau}{ S\left(1\right) -  I_0 \tau \left[  \left( 1 - e^{-\frac{t_i-t_{end}}{\tau}} \right) + \left( 1 - e^{-\frac{t_i}{\tau}} \right) \left( e^{-\frac{\delta_1}{\tau}} + e^{-\frac{\delta_2}{\tau}} \right) \right] -c + I_0 \tau} \\
  &=& \frac{I_0 \tau}{ S\left(1\right) -  I_0 \tau \left[  \left( 1 - e^{-\frac{t_i-t_{end}}{\tau}} \right) + \left( 1 - e^{-\frac{t_i}{\tau}} \right) \left( e^{-\frac{\delta_1}{\tau}} + e^{-\frac{\delta_2}{\tau}} \right) -1 \right] -c}
\label{f_eqn}
\end{eqnarray}

\subsection{Measurements}

\begin{figure*}[p!]
  \centering
  \begin{subfigure}[t]{0.45\textwidth}
	\includegraphics[width=\textwidth]{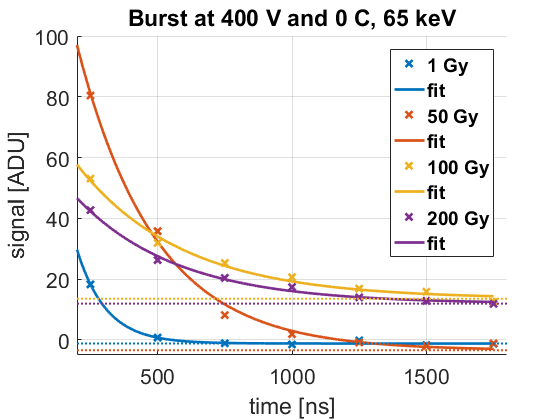}
	\caption{Average delayed signal in capacitors 2-8 (data points in time) at a bias voltage of 400~V for different accumulated doses and corresponding fits. Dotted lines indicate the offset level.}
	\label{fits}
  \end{subfigure}
\quad
  \begin{subfigure}[t]{0.45\textwidth}
	\includegraphics[width=\textwidth]{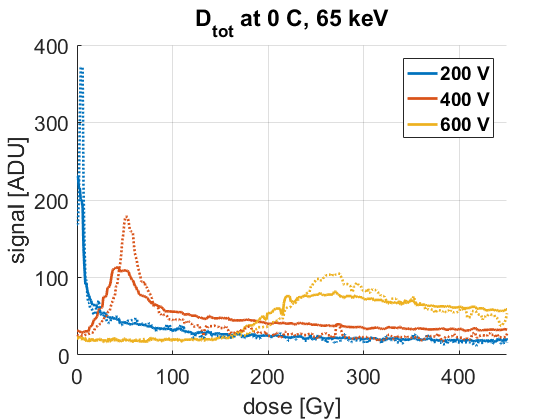}
	\caption{Total delayed response $D_{tot}$ as calculated by equation \ref{tot} for different bias voltages. The dotted lines represent a model independent estimate of $D_{tot}$ by summing the offset corrected signal of capacitors 2-8. }
	\label{d_tot}
  \end{subfigure}
\bigskip  
  \begin{subfigure}[t]{0.45\textwidth}
	\includegraphics[width=\textwidth]{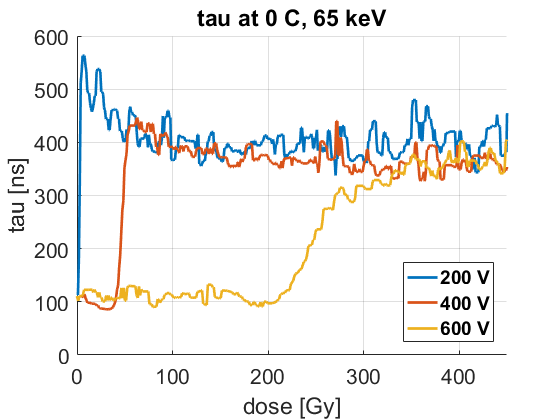}
	\caption{Time constant of the delayed response $\tau$ derived from the fit. }
	\label{tau}
  \end{subfigure}
\quad
  \begin{subfigure}[t]{0.45\textwidth}
	\includegraphics[width=\textwidth]{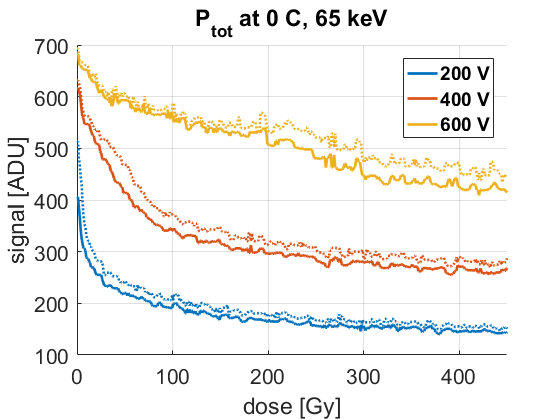}
	\caption{Total prompt response $P_{tot}$ as calculated by equation \ref{I_0} for different bias voltages. The dotted lines the offset corrected signal on the first capacitor $S(1)-c$ for comparison. }
	\label{p_tot}
  \end{subfigure}
\bigskip  
  \begin{subfigure}[t]{0.45\textwidth}
	\includegraphics[width=\textwidth]{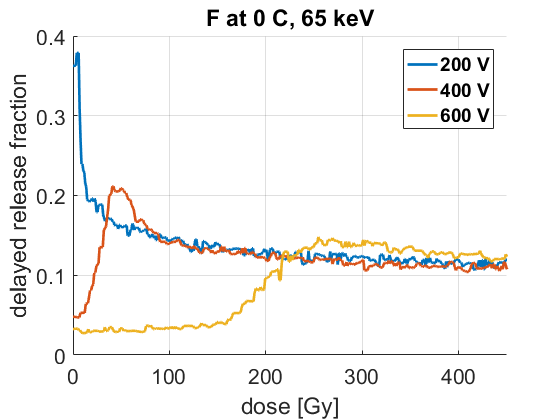}
	\caption{Fraction $F$ of the total response that is released delayed with respect to the prompt photocurrent as calculated by equation \ref{f_eqn} for different bias voltages. }
	\label{f_tot}
  \end{subfigure}
\quad
  \begin{subfigure}[t]{0.45\textwidth}
	\includegraphics[width=\textwidth]{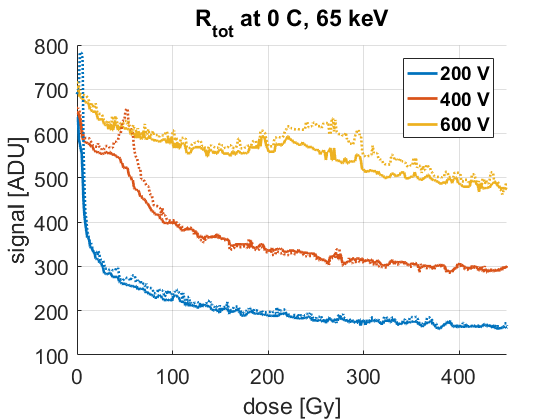}
	\caption{Total response $R_{tot}$ for different bias voltages. The dotted lines represent a model independent estimate of $R_{tot}$ by summing the offset corrected signals of all capacitors. }
	\label{r_tot}
  \end{subfigure}

  \caption{Measurements results in zone 4 at a temperature of 0~C in response to 65.2~keV photons.}
  \label{delay}
\end{figure*}

\begin{figure*}[tb!]
  \centering
	\includegraphics[width=0.8\textwidth]{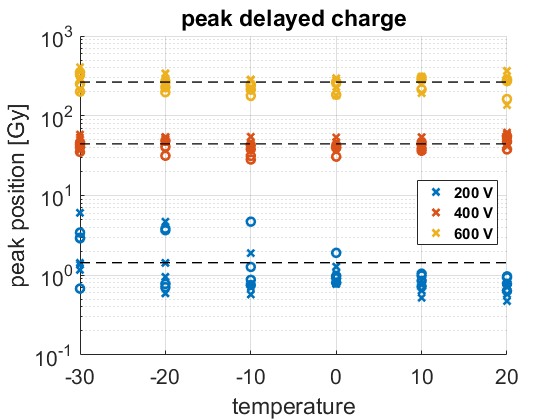}
	\caption{Dose for which the delayed charge reaches its maximum value as a function of temperature. Shown are results for 59.7~keV (`x') and 65.2~keV (`o') and all 4 zones. The dashed lines represent the averages for their respective bias voltage.}
	\label{pol}
\end{figure*}

In the absence of polarization the decay time of the delayed response is very short and all signal is collected within the first two integration intervals (capacitors 1 \& 2). However as polarization builds up in the sensor the signal collected in the `dark' period between turns of the electron bunches (capacitors 3 - 8) increases, see figure \ref{fits}. 

Correspondingly, the total delayed charge $D_{tot}$, the area under the fit curve in Figure \ref{fits}, changes with dose. As shown in Figure \ref{d_tot}, at first $D_{tot}$ increases with increased dose until it reaches a peak value before decreasing slowly again. This is less obvious at low voltages, as the peak value is reached almost immediately. 

The detrapping time, $\tau$, that is associated with the delayed release is shown in Figure \ref{tau}. It shows an almost bimodal distribution at higher bias voltages with a very sharp transition between the `unpolarized' value of approximately 100~ns and the value after polarization of about 400~ns. The observed detrapping times are well above the experimental uncertainty of about 10~ns, and are not observed when using a silicon sensor instead of the CdTe sensor. The transition between the two detrapping times occurs at approximately the same dose as the peak value in $D_{tot}$ (see Figure \ref{d_tot}). The nature of this transition remains unclear at this moment.

The fraction of the total charge, $F$, that is released delayed with respect to the prompt photocurrent and the total response are shown in Figures \ref{f_tot} and \ref{r_tot}. $F$ follows the same qualitative behavior as $D_{tot}$ and converges to a common value of 11-12\% for large doses\footnote{In this case the largest dose in the respective data set, between 212.5 and 550~Gy.}. For low doses and high voltages about 2-3\% of the total charge is released in a delayed fashion, which corresponds roughly to the amount of charge that is trapped during the prompt response. The total response $R_{tot}$ follows the characteristic pattern of polarization in CdTe reported previously \cite{static}. The total response exhibits an almost\footnote{For reasons unknown at this point the first 2--3 data points appear especially large.} flat plateau-like behavior before reaching a `knee' after which there is a rapid decline in response followed by a more gradual one.

The position of the peak in the delayed current also marks the onset of the decline of the total response. As shown in Figure \ref{pol}, within the investigated parameter ranges the amount of dose required to reach the peak does not depend appreciably on temperature, photon energy or dose rate, but only on the applied bias. We find that the peak is reached at 1.4~Gy~$\pm$~88\%, 44.4~Gy~$\pm$~17\% and 264~Gy~$\pm$~22\% at 200~V, 400~V and 600~V bias, respectively. The large spread of measurement values at 200~V bias is caused by the fact that at low bias voltages polarization happened almost immediately, within the first few data points, making a precise measurement difficult.

The position of the knee in Fig. \ref{r_tot} does not align well with the position of the knee when studied at a much lower dose rate of 11~Gy/min \cite{static}. At the lower dose rate the position of the knee is dependent on temperature and is located, in general, at a much higher dose value. For an operating temperature of 0~C these values are 110~Gy, 275~Gy, and 1430~Gy for voltages of 200~V, 400~V, and 600~V, respectively. We note that the lateral displacement effect described in \cite{static} does not explain our findings at high dose rates. Although its effects appear at roughly an order of magnitude less dose than the signal loss due to polarization. Lateral displacement would reduce the signal in the more heavily dosed regions of the detector, e.g., zone 4 in Figure \ref{zone_map}, by displacing it towards the less heavily dosed pixels surrounding the beam, e.g., zone 3. This would show as a dose rate dependence in the data, which we did not see. One possible explanation of this difference between high and low dose rate illumination could be an additional detrapping due to deeper traps on much longer time scales, e.g., milliseconds to seconds as reported in literature \cite{TCT3}. The contribution of these deeper traps would be collected during the longer integration times used in \cite{static}. In the given experiment the detrapping from these deeper traps would contribute to the constant offset $c$ in the measurements.

\begin{figure*}[tb!]
  \centering
  \begin{subfigure}[t]{0.45\textwidth}
	\includegraphics[width=\textwidth]{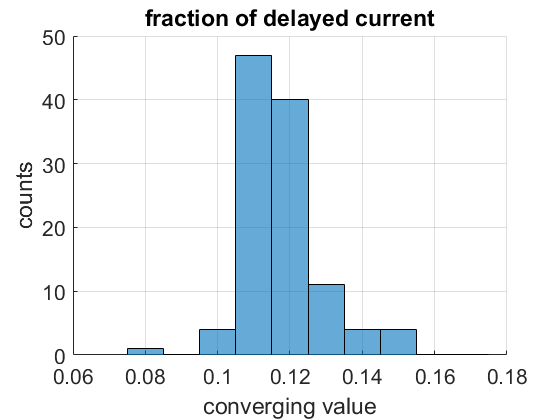}
	\caption{Histogram of the fraction of charge that is released delayed with respect to the prompt photocurrent for large doses. $F$ is independent of bias voltage and temperature and has an average value of 0.117~$\pm$~11\%.}
	\label{f_hist}
  \end{subfigure}
\quad
  \begin{subfigure}[t]{0.45\textwidth}
	\includegraphics[width=\textwidth]{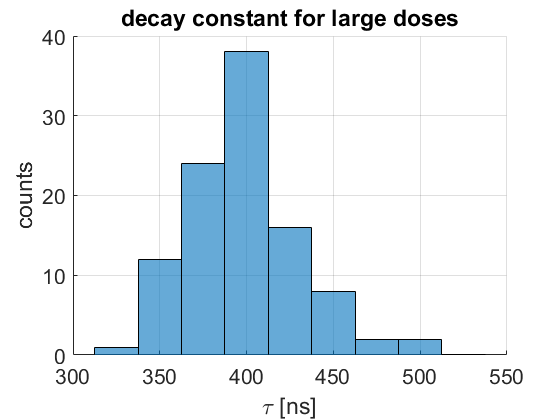}
	\caption{Histogram of the time constant associated with the delayed release of charge with respect to the prompt photocurrent for large doses. $\tau$ is independent of bias voltage and temperature and has an average value of 389~ns~$\pm$~13\%.}
	\label{tau_hist}
  \end{subfigure}
  
  \caption{Histograms of all measurement values for detrapping fraction $F$ and detrapping time $\tau$. For large doses both $F$ and $\tau$ are independent of energy, temperature, dose rate and bias voltage.}
  \label{hist}
\end{figure*}

However since we know that only a small, fixed amount of charge is released on those timescales (see section on basic testing), we conclude that the charge is lost due to recombination of charge carriers, which can reasonably be expected to be larger in presence of large currents.

For large doses all measurements converge to a similar value, $F$, of the fraction of charge that is released delayed with respect to the prompt photocurrent (see Figure \ref{f_hist}) and correspondingly to a similar detrapping time constant (see Figure \ref{tau_hist}). The average value of $F$ is 0.117~$\pm$~11\% and the average value of $\tau$ is 389~ns~$\pm$~13\%. When applying large biases we could also determine the average detrapping time before polarization set in (not shown), which is 139~ns~$\pm$~19\%.

\section{Conclusions}

CdTe of 750~$\upmu$m thickness, with its high quantum efficiency, expands the usable energy range for detector systems from about 25~keV ($\approx$20\% QE with a 500~$\upmu$m silicon sensor) to about 150~keV ($\approx$20\% QE with a 750~$\upmu$m CdTe sensor), with $>$90\% quantum efficiency to about 65~keV. 


Our characterization of the timing properties of CdTe sensors bonded to our previously developed ASICs has led us to the conclusion that the material is suitable for use in a broad range of experimental applications requiring timing down to the sub-microsecond level. Most of the effects described in this work can be explained using accepted drift and diffusion theories and simple model assumptions for detrapping. As shown herein, an understanding of these effects is important in order to understand the limitations of measurements taken with the investigted detectors. We note that both Keck PAD and MM-PAD detectors with CdTe sensors have already been used in a number of experiments at CHESS and the APS, and further experiments are planned for both detectors at both sources. 

We found that in our hole collecting Schottky material several shallow defects play an important role. We observed signal persistence with an exponential decay with timescales ranging from approximately 150~ms at -30~C to about 90~ms at +20~C. An Arrhenius fit of the time constant revealed an activation energy of 59.3~$\pm$~7.2~meV, which is consistent with defects that were observed in this kind of material before \cite{trap2}. The total magnitude of this persistence is determined by the number of filled traps, which is a function of the applied voltage. We were able to estimate that at the moment the shutter closes approximately 1.04--2.16~$\times$~10$^9$~cm$^{-3}$ traps are filled. The total charge released from these traps is equivalent to the charge collected from approximately 77--161~keV x-ray energy per pixel. The persistence due to this shallow trap is insufficient to explain signal loss observed on the microsecond time scale, where the signal loss is attributed to (trap-assisted) recombination, a non-linear process.

We found the charge collection time in hole-collecting mode to be less than 120~ns for biases of 600~V and above. We showed that with such a short charge collection time we were able to fully separate the signal of two consecutive pulses at the APS with 153~ns bunch spacing, i.e., 6.5~MHz repetition frequency, thus avoiding pile-up effects in the sensor. For hole trapping times of approximately 3~$\upmu$s, this leads to anticipated charge collection efficiencies exceeding 96\%. 

We found that polarization occurs quickly at high dose rates, in fact much quicker than expected from investigations at low dose rates. The prompt response to the photons decreases almost immediately, and a delayed current, possibly due to detrapping of trapped charge carriers from another, even shallower defect, is observed. A higher bias voltage delays the onset of polarization, but ultimately an equilibrium situation is reached. For large accumulated doses approximately 11.7\% of the total response is released in a delayed fashion with a time constant of 389~ns. This indicates a different type of mechanism is responsible for the delayed release of charge than for the persistence effect on millisecond time scales, which saturates. Within the investigated range the results for the delayed release of charge were independent of beam energy, dose rate, temperature and bias, indicating that this second defect is unlikely to be a bulk defect.

The reset procedure introduced in \cite{static} was employed and found to be effective at removing all the effects of polarization described here.

In summary, the Schottky type hole-collection CdTe studied here can be used for quantitative scientific hard x-ray imaging at time scales down to approximately 100~ns. However, precise x-ray measurements require significant attention to details: It is important that the sensor is reset prior to the measurement and that the sensor can be biased to sufficiently high voltage to delay polarization effects as long as possible. Even then, it is important to account for signal persistence, as it might mask or mimic a very weak signal following a large pulse.


\acknowledgments
The authors would like to thank Dominic Greiffenberg from the Paul Scherrer Institute in Switzerland for valuable discussions and input, Tom Krawcyzk from CHESS for his support during the beamtime at CHESS, the entire team of APS sector 35 for their help and support during the APS experiment, Wataru Inui at Acrorad, Co., Ltd for his contributions to the sensor design and production, and Klaus Harkonen, Limin Lin, and Konstantinos Spartiotis at Oy Ayad Ltd for the hybridization of the assemblies.

This research is based on research conducted at the Cornell High Energy Synchrotron Source (CHESS), which is supported by the U.S. National Science Foundation and the U.S. National Institutes of Health/National Institute of General Medical Sciences via NSF award DMR-1332208, and by awards for x-ray detector research to S.M.G. from the U.S. Department of Energy (Award DE-SC0016035), and the W. M. Keck Foundation. The MM- PAD concept was developed collaboratively by our detector group at Cornell University and Area Detector Systems Corporation (ADSC), Poway, CA, USA.

This publication is also based upon work performed at the Dynamic Compression Sector supported by the Department of Energy, National Nuclear Security Administration, under Award Number DE-NA0002442 and operated by Washington State University. This research used resources of the Advanced Photon Source, a U.S. Department of Energy (DOE) Office of Science User Facility operated for the DOE Office of Science by Argonne National Laboratory under Contract No. DE-AC02-06CH11357.






\nocite{*}
\bibliographystyle{aipnum-cp}%

\end{document}